\documentclass[floatfix,reprint,amsmath,amssymb,aps]{revtex4-2} % for journal
\usepackage{graphicx}% Include figure files
\usepackage{dcolumn}% Align table columns on decimal point
\usepackage{bm}% bold math
\usepackage[T1]{fontenc}

\newcommand\T{\rule{0pt}{2.6ex}}       % Top strut
\newcommand\B{\rule[-1.2ex]{0pt}{0pt}} % Bottom strut
\newcommand\Tspace[1]{\rule{0pt}{#1ex}}
\newcommand\Bspace[1]{\rule[-#1ex]{0pt}{0pt}}

\begin{document}
% \preprint{APS/123-QED}

\title{Measurement of the $\mathrm{^{25}Al(d,n)^{26}Si}$ reaction and impact \\ on the $\mathrm{^{25}Al(p,\gamma)^{26}Si}$ reaction rate}

\author{E. Temanson} %\email{}
\author{J. Baker}
\author{S. Kuvin} \altaffiliation[Current address: ]{P-3, Los Alamos National Laboratory, Los Alamos, NM, 87545, USA} 
\author{K. Hanselman} 
\author{G.~W.~McCann}
\author{L. T. Baby}
\author{A. Volya}
\author{P. H\"{o}flich}
\author{I. Wiedenh\"{o}ver}
\affiliation{Department of Physics, Florida State University, Tallahassee, Florida 32306, USA}

% \date{\today}
\date{May 27, 2023}
% ===============================================================
\begin{abstract}
The $\mathrm{^{25}Al(p,\gamma)^{26}Si}$ reaction is part of a reaction network with impact on the observed galactic $^{26}$Al abundance. A new determination of the proton strength of the lowest $\ell=0$ proton-resonance in $^{26}$Si is required to more precisely calculate the thermal reaction rate. To this end, the $\mathrm{^{25}Al(d,n)^{26}Si}$ proton-transfer reaction is measured in inverse kinematics using an in-flight radioactive beam at the \texttt{RESOLUT} facility. Excitation energies of the lowest $^{26}$Si proton resonances are measured and cross sections are determined for the lowest $\ell=0$ resonance associated with the $3^{+}_{3}$ state at 5.92(2)~MeV. Coupled reaction channels (CRC) calculations using \texttt{FRESCO} are performed to extract the $\ell=0$ spectroscopic factor for the $3^{+}_{3}$ state. The proton width for the $3^{+}_{3}$ state in $^{26}$Si is determined to be $\Gamma_{p}$=2.19(45)~eV and the $(p,\gamma)$ resonance strength for the $3^{+}_{3}$ state is extracted as 26(10)~meV. This resonance dominates the $\mathrm{^{25}Al(p,\gamma)^{26}Si}$ reaction rate above 0.2~GK. 
\end{abstract}
\maketitle

% ===============================================================
\section{\label{sec:Intro} Introduction}

The observation of the 1.809~MeV $\gamma$-ray associated with the radioactive decay of $^{26}$Al in 1982 by \texttt{HEA0-3} marked the first experimental evidence of ongoing nucleosynthesis in the galaxy  \cite{Mahoney1982}. Further improvements in $\gamma$--ray astronomy have allowed telescopes such as \texttt{COMPTEL} \cite{Diehl1994} and \texttt{INTEGRAL} \cite{Bouchet2015} to probe the spatial emission of the 1.809~MeV $\gamma$-ray across the galactic plane. Based on the emissions along the plane of the galaxy and an increased intensity near known massive star groups, the likely production cites for $^{26}$Al are core-collapse supernova and Wolf-Rayet stars \cite{Diehl2021}.

The presence of a low-lying isomeric $0^+$ state, $^{26}$Al$^m$, whose population can only slowly reach thermal equilibrium with the $5^{+}$ $^{26}$Al$^g$ ground state, severely complicates the calibration of its nucleosynthesis. In discussions of the subject, beginning with Ref.\ \cite{Ward1980} it is found that thermal equilibrium between $^{26}$Al$^g$ and $^{26}$Al$^m$ will typically be reached at temperatures above 0.4 GK, where indirect $\gamma$--absorption and emission competes with the $T_{1/2}=6.345s$ $\beta^+$--decay of $^{26}$Al$^m$. Since this super-allowed $\beta^+$--decay exclusively populates the $^{26}$Mg ground state, it is in effect bypassing the 1.808~MeV $\gamma$--ray, on which the $\gamma$--ray astronomical observation is based. In consequence, the nucleosynthesis yield of $^{26}$Al$^g$ depends on the nuclear reactions leading to both $^{26}$Al$^g$ and $^{26}$Al$^m$, their destruction rates, and the rates of equilibration processes between them. Both the ground and isomeric states are produced by the reaction $\mathrm{^{25}Mg(p,\gamma)^{26}Al^{g,m}} $. However, at sufficiently high temperatures, the $\mathrm{^{24}Mg(p,\gamma)^{25}Al}$ reaction followed by the $\mathrm{^{25}Al(p,\gamma)^{26}Si}$ reaction becomes competitive to produce $^{26}$Si, which then decays exclusively to $^{26}$Al$^m$. 

Motivated by the implications on $\gamma$-ray astronomy, multiple works have addressed the spectrum of proton resonances in $^{26}$Si and their relative impact on the $\mathrm{^{25}Al(p,\gamma)}$ reaction. Several experiments used indirect methods, through the $^{28}$Si(p,t) \cite{Bardayan2006, Matic2010}, the $^{24}\mathrm{Mg}(^3\mathrm{He},n)$ \cite{Parpottas2004} and the $^{29}\mathrm{Si}(^3\mathrm{He},^6\mathrm{He})$ \cite{Caggiano2002} reactions. Other works used $\gamma$--spectroscopy after the ($^3$He,n) reaction, \cite{DeSereville2010, Perello2022} to establish some level energies with high accuracy. A comprehensive analysis of the resonance spectrum by Chipps {\it et al.} \cite{Chipps2016} combined the available information from multiple works and, in particular, identified the dominant $\ell=0$ resonance with the 3$^+_3$ state at 5.9276(10)~MeV of excitation energy. In the following, we will use the excitation energy from the NNDC data base \cite{nndc}, 5.9294(8)~MeV and the corresponding center-of-mass resonance energy 0.4154(8)~MeV.   

Information on the strength of the $^{26}$Si resonances in question was also obtained in studies investigating the isospin-mirror nucleus $^{26}$Mg via the $\mathrm{^{25}Mg(d,p)^{26}Mg}$ reaction \cite{Hamill2020,Burlein1984,Arciszewski1984}. The most recent study of $^{26}$Mg by C. B. Hamill {\it et al.} \cite{Hamill2020} focused on the important 3$^+$ and 0$^+$ analog states at 6.125 and 6.255~MeV respectively and reported an upper limit for the lowest 1$^+$ state at 5.69 MeV. 

As a first radioactive-beam experiment at FSU's {\sc resolut} facility \cite{resolut2013}, Peplowski {\it et al.} \cite{Peplowski2009} measured the proton-spectra following the $\mathrm{^{25}Al(d,n)^{26}Si}$ transfer reaction in inverse kinematics and extracted the proton width of 19--52 meV for 0.4154(8)~MeV resonance. Subsequently Bennett {\it et al.} \cite{Bennett2013} and Liang {\it et al.} \cite{Liang2020} used the $\beta^+$--decay of $^{26}$P to populate the resonance spectrum in $^{26}$Si and to determine the $\gamma$--proton branching ratio of the 0.4154(8)~MeV resonance. With the same goal, Perello {\it et al.} \cite{Perello2022} populated the resonances through the $^{24}$Mg($^{3}$He,n)$^{26}$Si reaction and performed neutron-$\gamma$ coincidence spectroscopy, obtaining results consistent with \cite{Bennett2013} and \cite{Liang2020}. Those works determined the partial $\gamma$-width of the 0.4154(8)~MeV resonance by combining their respective $\gamma$--proton branching ratios with the partial proton width extracted by Peplowski {\it et al.} \cite{Peplowski2009}. 

% \vfill\eject % column break
The present work describes a re-measurement of the $\mathrm{^{25}Al(d,n)^{26}Si}$ transfer reaction to determine the $\ell=0$ proton-strength of the 3$^+_3$ resonant state with improved methods and higher precision than Peplowski {\it et al.} \cite{Peplowski2009}.

% ===============================================================
\section{\label{sec:Exp} The $\mathrm{^{25}Al(d,n)^{26}Si}$ experiment}

The experiment used a beam of the radionuclide $^{25}$Al, produced with the {\sc resolut} facility \cite{resolut2013} at the John D. Fox accelerator laboratory of Florida State University. The beam of $^{25}$Al was produced by bombarding a gas-cell target filled with D$_2$ gas, cooled to liquid-nitrogen temperatures, with the primary beam of $^{24}$Mg, accelerated by the Tandem and superconducting LINAC facility to 142 MeV. The reaction products from the $\mathrm{^{24}Mg(d,n)^{25}Al}$ reaction were separated and focused onto a secondary target through the {\sc resolut} facility, using a superconducting resonator to actively sharpen the particle energies and a magnetic spectrometer to separate the desired reaction products from the primary beam. A description of the principle of operation, the beam optics and a list of radioactive beams used in experiments is given in \cite{resolut2013}. The optimal beam intensity was obtained at a magnetic rigidity of 0.5597 Tm, corresponding to 102 MeV of $^{25}$Al in the 13$^+$ charge state.  

\begin{figure}[!htbp]
    \centering
    \includegraphics[width=\columnwidth]{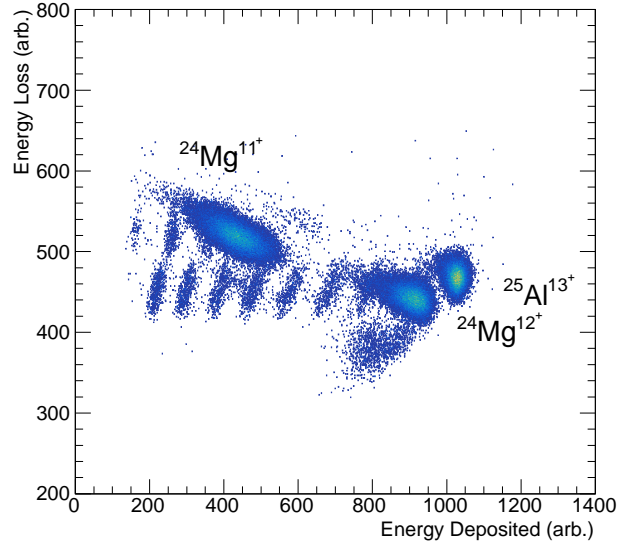}
    \caption{Secondary beam composition measured in the gas ionization chamber. }
    \label{fig:BeamProfile}
\end{figure}

The beam intensity and purity were monitored continuously in a zero-degree ion chamber, which is part of the experimental setup described below. The beam components were identified and separated by their characteristic energy losses, as displayed in Fig.~\ref{fig:BeamProfile}, where the $^{25}$Al particles were well-separated from the $^{24}$Mg beam background. The secondary beam of interest, $^{25}$Al, had an average beam intensity of $7.5 \times 10^{3}$ particles per second, 25\% of the total number of particles reaching the secondary target. The $^{25}$Al beam particles were also identified and and selected during the event analysis through their characteristic time-correlation with the accelerator RF reference. 

The secondary beam of $^{25}$Al bombarded a 512 $\mu g/cm^2$ thick deuterated polyethylene ($CD_{2}$) film. For detection of the decay protons, two annular Micron Semiconductor S2-type detectors with 64 $\mu m$ and 1000 $\mu m$ thicknesses were positioned at 69 mm and 82 mm downstream of the target, subtending angles between 8$^\circ$ and 22$^\circ$ from the beam axis. Each annular double-sided strip detector was read out in 16$\times$16 channels, allowing for extraction of $\theta$ and $\phi$ for each particle. Signals from the beam particles and reaction residues were detected in a position-sensitive ion-chamber filled with 35 Torr of Isobutane gas. The ion chamber, see Ref.\ \cite{Lai2018}, was segmented into four depth regions. The first two sections of 40 mm depth allowed for a position-determination of the incoming particle in horizontal and vertical directions, the third and fourth section of 80~mm and 200~mm depth were used to determine the differential energy-loss and residual energy signals. 
% \vfill\eject % used for column break

\begin{figure}[!htbp]
\centering
    \includegraphics[width=\columnwidth]{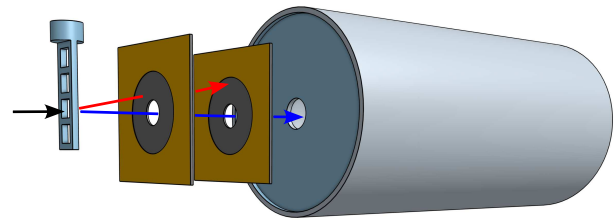}
    \caption{Schematic representation of the experimental setup including the target ladder, silicon telescope and position sensitive ion chamber. }
\label{fig:assemly}
\end{figure}

% \vfill\eject % column break
As the main trigger condition, events in either of the silicon detectors were recorded, along with the signals from the heavy-ion reaction residues and the timing information relative to the accelerator RF reference. In addition to the silicon-triggered events, the beam composition was continuously sampled by recording one of every 1000 events triggering the ion chamber. The silicon detectors were energy calibrated using standard calibration sources. The ion chamber sections were calibrated by matching the signals of the beam-components, whose energy was determined by the rigidity-measurement of {\sc resolut}, to a calculation of the deposited energies with the program \verb|CATIMA| \cite{CAtima}. 

\begin{figure}[!htbp]
    \centering
    \includegraphics[width=\columnwidth]{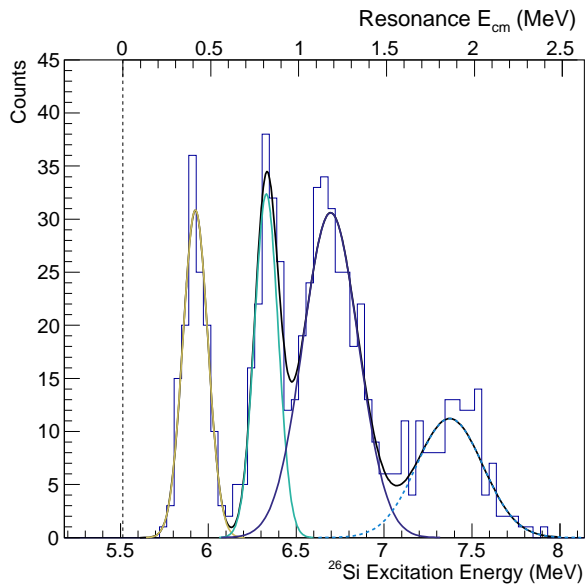} 
    \caption{Excitation energy spectrum of $^{26}$Si. The secondary x-axis represents the associated center-of-mass proton resonance energies. The state of interest for astrophysical environments is the $3^{+}_{3}$ at 5.92 MeV. } 
    \label{fig:ExcEnergy}
\end{figure}

% ===============================================================
\subsection{\label{sec:pSpectrum} Experiment analysis}

The excitation energy of $^{26}$Si was reconstructed through the $^{26}\textrm{Si}^{*}$ $\rightarrow$ $^{25}\textrm{Al} + p$ decay path where both particles were detected in coincidence. The protons and $^{25}$Al output channels were identified using their characteristic energy-loss vs. energy correlations observed in the silicon and ion-chamber detectors, respectively. Additional gates on the coincidence timing and a geometric correlation between both DSSD silicon detectors are applied. From the energies and emission directions of the coincident proton and $^{25}$Al particles the excitation energy of $^{26}$Si states was reconstructed through an invariant mass analysis. The resulting spectrum is displayed in Fig. \ref{fig:ExcEnergy}, showing a strong peak at 5.92 MeV, a second peak at 6.33 MeV, a wider structure at 6.70 MeV and higher excited states near 7.40 MeV. The 5.92 MeV peak is identified with the 5.9294(8) MeV 3$^+_3$ state \cite{nndc}, at a resonance energy of 0.4154(8) MeV, the highest-impact resonance for the $\mathrm{^{25}Al(p,\gamma)^{26}Si}$ reaction. 

A Monte-Carlo simulation was used to determine the coincident-detection efficiency of protons and reaction residues. The kinematics of the $(d,n)$ neutrons and the subsequent proton decays were simulated in correlation, where the neutron angular distribution was weighted with a typical c.m. angular distribution, while all c.m.-proton angles entered evenly. 

% ===============================================================
\begin{table}[!htbp] %htbp
\centering
\caption{$^{26}$Si excited states observed in $^{26}$Si through the $^{25}$Al(d,n)$^{26}$Si reaction, with the corresponding cross sections.}
\begin{ruledtabular}
\begin{tabular}{ccccc}
%------------------------------------------- Title
\begin{tabular}[c]{@{}c@{}} E$_x$ \T\B\\ (MeV) \end{tabular} &
\begin{tabular}[c]{@{}c@{}} Adopted E$_x$\footnote[1]{Nuclear data reference \cite{nndc,AME2020}} \T\B\\ (MeV) \end{tabular} &
\begin{tabular}[c]{@{}c@{}} E$^{c.m.}_R$ \footnotemark[1] \T\B\\ (MeV) \end{tabular} &
$J^\pi$ & 
\begin{tabular}[c]{@{}c@{}} $\sigma^{stat.}_{syst.}$ \T\B\\ (mb) \end{tabular} \\ \hline

%------------------------------------------- 3+
5.92(2) & 5.9294(8) & 0.4154(8) & $3^{+}_{3}$ & $5.83^{\pm 0.09}_{\pm 0.78}$ \Tspace{2.6}\Bspace{1.2} \\

%------------------------------------------- 2+
6.33(2) &  
    \begin{tabular}[c]{@{}l@{}} 6.2953(24) \T\B\\ 6.3827(29) \end{tabular} &
    \begin{tabular}[c]{@{}l@{}} 0.7813(24) \T\B\\ 0.8687(29) \end{tabular} &
    \begin{tabular}[c]{@{}l@{}} $2^{+}_{6}$ \T\B\\ $2^{+}_{7}$ \end{tabular} & 
$10.02^{\pm 0.09}_{\pm 1.42}$ \Tspace{4.0}\Bspace{4.0} \\

%------------------------------------------- 3-
6.70(2) & 6.787(4) & 1.273(4) & $3^{-}_{1}$ & $30.12^{\pm 0.06}_{\pm 4.26}$ \T\B

\end{tabular}
\end{ruledtabular}
\label{table:Si26CrossSection}
\end{table}
% ===============================================================

The cross sections listed in Table~\ref{table:Si26CrossSection} were extracted from the peak areas in the spectrum, the integrated number of beam particles determined from the ion chamber data, the target density and the simulated detection efficiency for the respective excitation energy. Here, the resonances are assumed to decay exclusively by proton emission. The uncertainties in the cross sections are dominated by  systematic uncertainties in the efficiency calculations and target thickness. 

% \vfill\eject % column break
The measured cross section of the 3$^+$ state agrees with the previous measurement by Peplowski~{\it et al.} \cite{Peplowski2009}, 8.7$\pm$3~mbarn, with improved statistics and smaller uncertainty. Two additional peaks at 6.3~MeV and 6.7~MeV were measured as well. Peak 2 in Fig.~\ref{fig:ExcEnergy} is comprised of two states with a combined cross section of 10(2)~mb. Peak 3, observed with a total cross section of 30(4)~mb, is likely associated with the 6.787~MeV 3$^-_1$ state, but it is unclear if there is any contribution from additional states such as 6.76 or 6.81~MeV reported in Ref.~\cite{nndc}.

\begin{figure}[!htbp]
    \centering
    \includegraphics[width=0.95\columnwidth]{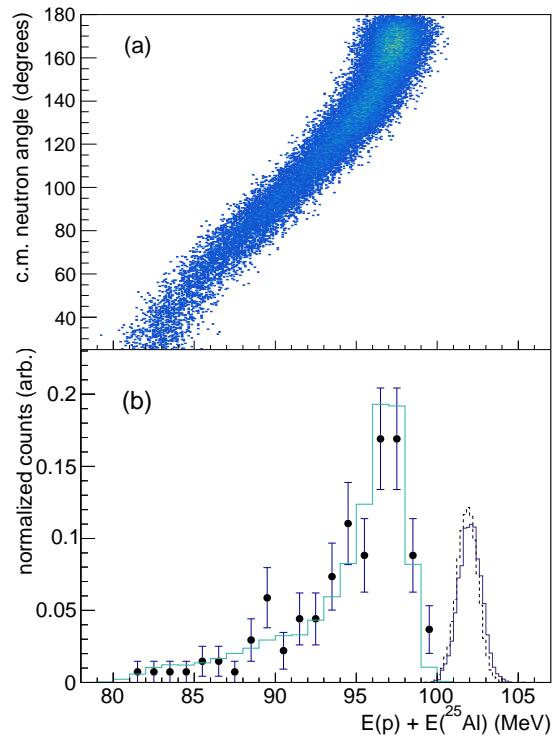}
    \caption{ (Panel a) Correlation between the $^{25}$Al+p sum-energy and neutron-emission angle extracted from the Monte-Carlo simulation based on the (d,n) angular distribution from CRC theory (see text).  (Panel b) The 5.92~MeV $3^{+}_{3}$ experimental $^{25}$Al+p sum-energy distribution compared to the results of the Monte-Carlo simulation. Shown for comparison is the experimental and simulated $^{25}$Al beam energy profile. }
    \label{fig:FragDistribution}
\end{figure}

While the neutrons emitted in the (d,n) reaction remain undetected in the current experiment, the neutron energies leave an imprint as ``missing'' energy in the total energy of the $^{26}$Si compound nucleus, which is reconstructed from the proton and $^{25}$Al final-state particles. In Panel(a) of Fig.~\ref{fig:FragDistribution}, the simulated correlation between the neutron angle in the center-of-mass system and the final-state $^{25}$Al+p sum-energy is displayed, based on calculated neutron angular distribution from the CRC model for this reaction, which itself will be described in the following section. Because of the strong kinematic variation of the (d,n) neutron energy with the emission angle, the energy distribution also contains information on the angular distribution. The distribution of the experimental $^{25}$Al+p sum-energies for the 5.92~MeV 3$^+$ state is displayed in Panel(b) of Fig.~\ref{fig:FragDistribution}.  The simulated energy spectrum shows an excellent fit to the data and the quality of agreement argues for the applicability of the CRC reaction model in the experimental analysis. A similar analysis has also been described and applied in Ref.\ \cite{Belarge2016}. 

% ===============================================================
\subsection{Discussion of a potential $^{26}$Si state at 5.95~MeV.} 
A recent paper by Canete~{\it et al.} \cite{Canete2021} described a $1^{-}$ state in $^{26}$Mg located at 5.710 MeV, which was assigned as a mirror-state to a potential 5.95~MeV state in $^{26}$Si, creating a large uncertainty in the nucleosynthesis calculations. This state had been observed by Caggiano {\it et al.} \cite{Caggiano2002} through a ($^3$He,$^6$He) reaction at a magnetic spectrograph and subsequently confirmed by Parpottas {\it et al.} \cite{Parpottas2004} in a ($^3$He,n) reaction, which was analyzed through neutron time-of-flight spectroscopy. Applying the parameters suggested by Canete {\it et al.}, the single-proton strength of this resonance would be small and below the sensitivity of our experiment. 

However, we observe that the evidence for the existence of a resonance at this energy is weak: The $\mathrm{^{24}Mg(^{3}He,n\gamma)^{26}Si}$ reaction measured by Perello~{\it et al.} \cite{Perello2022} was performed with the same reaction at the same energy as Parpottas~{\it et al.} and observed two resonances at 0.375(2)~MeV (0$^+$) and 0.4138(11)~MeV (3$^+$), consistent with the number of peaks observed by Parpottas~{\it et al.} in this region, but each at slightly lower excitation energies.  It can therefore be assumed that the energy calibration of Parpottas~{\it et al.} was slightly wrong. The Caggiano ~{\it et al.} experiment is now the only information pointing to an additional state at this energy and has not been independently verified. It also seems unlikely that such a 1$^-$ state would not have been observed by Perello~{\it et al.} through its expectedly dominant $\gamma$--decay. We conclude that there is insufficient evidence for this 1$^-$ resonance in $^{26}$Si and therefore do not include in the following analysis.   

% ===============================================================
\subsection{\label{sec:3p} $3^{+}_{3}$ Resonance Parameters}

The main goal in the analysis is the determination of the resonance proton-width for the 3$^+$ resonance, which is closely related to its $\ell=0$ spectroscopic factor. This experiment determined the total cross section for the population of this state in the $(d,n)$ reaction, but the limited experimental sensitivity to the neutron-angular distribution does not allow for the disentanglement of the $\ell=0$ and $\ell=2$ contributions. Here, we are relying on the mirror-reaction to guide the extraction of the relative $\ell=0$ component. 

The reaction mechanism was investigated using a coupled reaction channels (CRC) calculation using the program \texttt{FRESCO} \cite{Fresco}. The entrance channel utilized the global deuteron optical model potentials from Haixia An and Chonghai Cai \cite{Haixia2006}. The exit channel was calculated using the global optical model parameters from A.J. Koning and J.P. Delarouche \cite{Koning2003} assuming a neutron energy of 300 keV. The final states of interest in $^{26}$Si are proton-unbound and were calculated using a weak-binding approximation of 50 keV, which is justified through the relatively small widths of the states in question. 

\begin{figure}%[!htbp]
    \centering
    \includegraphics[width=\columnwidth]{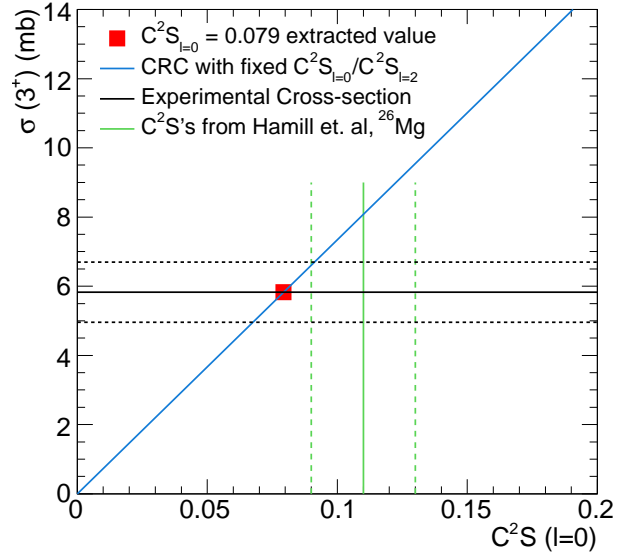} 
    \caption{Reaction cross sections from the CRC calculation (see text) in relation to the $C^{2}S_{l=0}$ spectroscopic factor. The values and error limits for the experimental cross section and the mirror-nucleus spectroscopic factors by Hamill {\it et al.} \cite{Hamill2020} are marked. The blue line represents the cross section calculated while scaling the $C^{2}S_{l=0}$ and $C^2S_{l=2}$ proportionally. }
    \label{fig:C2S_Sensitivity}
\end{figure}

In Fig.~\ref{fig:C2S_Sensitivity} the method for extraction of the $\ell=0$ spectroscopic factor for the 3$^{+}$ resonance is illustrated. The experimental cross section and error interval are marked on the vertical axis. The spectroscopic factors determined in the mirror nucleus $^{26}$Mg by Hamill {\it et al.} \cite{Hamill2020} and the corresponding CRC cross section are also shown. Applying the mirror-nucleus spectroscopic factors directly would predict a slightly higher cross section than measured. Scaling the spectroscopic factors of Hamill {\it et al.} down to reproduce the observed cross section, arrives at the extracted $\ell=0$ spectroscopic factor $C^{2}S_{l=0} = 0.079(16)$. 

% \vfill\eject % column break
In order to extract the proton-resonance width from the $\ell=0$ spectroscopic factor, we employed the R-matrix expression
\begin{equation}
\Gamma_{p}=C^{2}S_{l=0} \ \Gamma_{s.p.} = C^{2}S_{l=0} \ 
\frac{\hbar^2  P_{c} }{\mu~r_{c}} u^2(r_{c}) 
\label{eq:singe-particle-width}
,
\end{equation}

where $r_{c}$ is the channel (interaction) radius, $\mu$ is the reduced mass, $P_{c}$ is the penetrability of the Coulomb and centrifugal barriers and $u(r_{c})$ is the radial wave function, which was extracted from the CRC calculation, evaluated at the channel radius $r_{c}=r_{0}(A^{1/3} + 1)$ with $r_{0}=1.25$. At the adopted resonance energy, the proton single-particle width was calculated at the experimental resonance energy, resulting in $\Gamma_{s.p.} = 27.6(14)$~eV. 

The extracted $C^{2}S_{l=0}$ was multiplied with the proton single-particle width to extract $\Gamma_{p} = 2.19(45)$~eV for the 3$^+$ resonance, which is in agreement with the previous value extracted by Peplowski~{\it et al.}, within error limits \cite{Peplowski2009}. The error associated with $\Gamma_{p}$ is largely due to the systematic uncertainty in determining the $\ell=0$ spectroscopic factor, both the statistical and systematic uncertainty are reported in Table~\ref{table:ResoParam}. 

% ===============================================================
\section{\label{sec:astro} Thermal reaction rates }

% ===============================================================
\begin{table*}[!htbp]
\centering
\caption{Parameters of the most important proton resonances in $^{26}$Si with reference sources. The values of the underlined references were used in the reaction rate calculations (See  Fig.~\ref{fig:RRate} and Fig.~\ref{fig:RRateDifference}) }
\begin{ruledtabular}
\begin{tabular}{cllccccc}

% ================================================
$J^\pi$ &  Reference & E$^{c.m.}_R$ (MeV) &  $C^{2}S$ & $\Gamma_p$ (eV) & $\Gamma_\gamma/\Gamma_p$ & $\Gamma_\gamma$ (eV) & $\omega\gamma$ (meV)  \\ \hline
% ------------------------------------ 1+

$1^{+}_{1}$ & \underline{Hamill \textit{et al.}} \cite{Hamill2020} & $0.1622(3)$\footnote[1]{Nuclear data references \cite{nndc,AME2020}} & $< 5.70\times10^{-3}$ & $< 8.90\times10^{-9}$ & & $0.12$\footnote[2]{Shell model calculation by Richter {\it et al.} \cite{Richter2011}} & $< 2.60\times10^{-6}$  \T\B \\

\hline % ------------------------------------ 0+

$0^{+}_{4}$ &  Hamill \textit{et al.} \cite{Hamill2020} & 0.3761(3)\footnotemark[1] & 0.042(10) & 0.0042 & & 0.0088\footnotemark[2] & 0.24 \T\B \\

$0^{+}_{4}$ &  \underline{Perello \textit{et al.}} \cite{Perello2022} & 0.375(2) &   & 0.0042 & & 0.0075\footnote[3]{Updated Richter {\it et al.}'s \cite{Richter2011} result using new location of the $0^{+}_{4}$} & 0.22 \T\B \\

\hline % ------------------------------------ 3+

$3^{+}_{3}$ &  Bennett \textit{et al.} \cite{Bennett2013} & 0.4149(15) &    & 2.9(10)\footnote[4]{Adopted from Peplowski {\it et al.} \cite{Peplowski2009}} & 0.014(9)\footnote[5]{A complimentary reanalysis by P\'erez-Loureiro \textit{et al.} \cite{Perez2016} reported $\Gamma_\gamma/\Gamma_p$=0.015(5)} & $0.040(30)$ & $23(17)$ \T\B \\

$3^{+}_{3}$ &  Hamill \textit{et al.} \cite{Hamill2020} & 0.4154(8)\footnotemark[1] & 0.11(2), 0.27(6) & 2.9(10)\footnotemark[4] & 0.014(9) & 0.040\footnote[7]{No uncertainties were provided in this reference.} & 23\footnotemark[7] \T\B \\

$3^{+}_{3}$ &  Liang \textit{et al.} \cite{Liang2020} & 0.4124(19)\footnote[6]{Error-weighted mean value from \cite{Liang2020,Janiak2017,Thomas2004}} &  & 2.9(10)\footnotemark[4] & 0.0207(75) & 0.060(30) & 34(17) \T\B \\

$3^{+}_{3}$ &  Perello \textit{et al.} \cite{Perello2022} & 0.4138(11) &  & 2.9(10)\footnotemark[4] & 0.025(14) & 0.071(32) & 40(17) \T\B \\

$3^{+}_{3}$ & \underline{This work} & 0.4154(8)\footnotemark[1] & 0.079(16), 0.20(4) & $2.19^{\ \pm 0.04\ stat.}_{\ \pm 0.42\ syst.}$ \Tspace{2.6}\Bspace{1.2} & 0.0207(75)$^{lit.}$ & 0.045(17) & 26(10) \T\B \\
% ================================================
\end{tabular}
\end{ruledtabular}
\label{table:ResoParam}
\end{table*}
% ===============================================================

Table~\ref{table:ResoParam} summarizes the relevant proton resonance parameters in $^{26}$Si. Included alongside this work's updated $3^{+}_{3}$ parameters are the adopted $1^{+}_{1}$ and $0^{+}_{4}$ parameters from the $\mathrm{^{25}Mg(d,p)^{26}Mg}$ reaction by Hamill~{\it et al.} \cite{Hamill2020} and the  $\mathrm{^{24}Mg(^3He,n)^{26}Si}$ reaction by Perello~{\it et al.} \cite{Perello2022}. The Maxwell-averaged two-body reaction rate was calculated using the Breit-Wigner approximation for narrow resonances

\begin{equation}
N_{A}\langle\sigma\nu\rangle_{r}=\frac{1.5399\times10^{11}}{(\mu T_{9})^{3/2}}
    \sum\limits_{i} (\omega\gamma)_{i} \: e^{-11.605 E^{c.m.}_{Ri}/T_{9}} ,
\end{equation}

where $\mu$ is the reduced atomic mass in $a.m.u.$ from the AME 2020 compilation \cite{AME2020}, E$^{c.m.}_{Ri}$ is the center of mass resonance energy in MeV, $T_{9}$ is the temperature in units of GK. The resonance strength, $(\omega\gamma)$, is defined as

\begin{equation}
    \omega\gamma = \frac{2J_{r} + 1}{(2J_{p} + 1)(2J_{T}+1)} \left( \frac{\Gamma_{p}\Gamma_{\gamma}}{\Gamma_{p}+\Gamma_{\gamma}} \right)
    \label{eq:res-strength}, 
\end{equation}

where the spin of the proton is $J_{p}=1/2$, the spin of the $^{25}$Al ground state is $J_{T}=5/2$ and $J_{r}$ is the spin of the resonance. The proton partial width derived in this work is $\Gamma_{p}$=2.19(45) for the $3^{+}_{3}$ state, as discussed in Sec.~\ref{sec:3p}. Because the 3$^+$ resonance decays predominantly by proton emission, the partial $\gamma$--width becomes the determining factor of the resonance strength $\omega\gamma$. However, this partial $\gamma$--width can be determined now with higher precision using the $\gamma$--proton branching ratios determined by other recent experiments \cite{Bennett2013,Liang2020,Perello2022}. A summary of the previous results is represented in  Table~\ref{table:ResoParam}. 

An interesting difference between the results obtained by Bennett {\it et al.} \cite{Bennett2013} and Liang {\it et al.} \cite{Liang2020} lies in the respective normalization of the observed $\beta^+$--delayed $\gamma$--decays relative to the $\beta^+$--delayed proton decays. Bennett {\it et al.} only measured the $\beta$--delayed $\gamma$--decay intensity from the $3^{+}_{3}$ state and normalized it relative to the $\beta$--delayed proton emission intensity from Thomas {\it et al.} \cite{Thomas2004}. In contrast, Liang {\it et al.} measured both branches in a single experiment. The proton-decay strength of 17.96(90)~$\%$ from Thomas {\it et al.} \cite{Thomas2004} also appears problematic, since their values are inconsistent with the subsequent experiments by Liang {\it et al.} (11.1(12)~$\%$) \cite{Liang2020} and Janiak {\it et al.}'s (10.4(9)-13.8(10)~$\%$) \cite{Janiak2017}, which are consistent with each other. 

It should be noted that the $\beta^{-}$--delayed $\gamma$--ray intensity for the 1742~keV $3^{+}_{3} \rightarrow 3^{+}_{2}$ transition measured by Liang {\it et al.} \cite{Liang2020} had significant statistical uncertainty and the authors used an error-weighted mean combining their results, Bennett {\it et al.} \cite{Bennett2013} and P\'erez-Loureiro \textit{et al.} \cite{Perez2016}, then calculating the proton-branching ratio from the Liang {\it et al.} proton-decay probability. 

The resonance parameters for the $1^{+}_{1}$ and $0^{+}_{4}$ states were adopted from the isospin-mirror reaction performed by Hamill~{\it et al.} \cite{Hamill2020}, the shell model calculations by Richter~{\it et al.} \cite{Richter2011} and the $\mathrm{^{24}Mg(^{3}He,n\gamma)^{26}Si}$ reaction by Perello~{\it et al.} \cite{Perello2022}. A 30\% error was assumed for the resonance strength of these two states in the calculation of the reaction rate uncertainties. 

The direct-capture component of the reaction rate is approximated using 
\begin{multline}
N_{A}\langle\sigma\nu\rangle_{dc}=
    7.8327\times10^{9} \:(S_{eff})\: \left( \frac{Z_{T}}{\mu T^{2}_{9}} \right)^{1/3} \\
    \times \exp\left(-4.2487 \left( \frac{Z^{2}_{T}\mu}{T_{9}} \right)^{1/3} \right)
\end{multline} 
and
\begin{equation}
    S_{eff}\approx S(0) \left( 1 + 0.09807 \left( \frac{T_{9}}{Z^{2}_{9}\mu} \right)^{1/3} \right),
\end{equation}

where $Z_{T}$ is the atomic number of $^{25}$Al and $S_{eff}$ is the effective astrophysical S-factor. 
The zero energy point of the S-factor, S(0), which is dominated by the effects of direct proton capture, was adopted from Matic~{\it et al.} \cite{Matic2010} and the value 0.028 MeV-b, with an estimated 30$\%$ uncertainty.

\begin{figure}%[!htbp]
    \centering
    \includegraphics[width=\columnwidth]{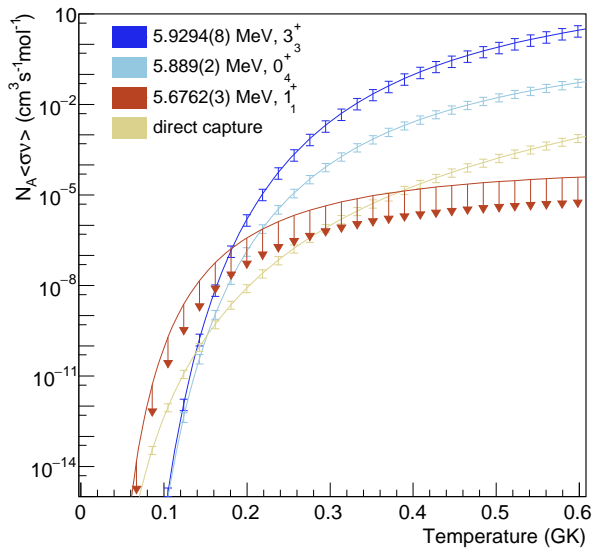}
    \caption{Thermal rate for the $\mathrm{^{25}Al(p,\gamma)^{26}Si}$ reaction. The total rate is separated into direct capture (gold) and three individual resonance contributions, $1^{+}_{1}$ (red), $0^{+}_{4}$ (light blue) and $3^{+}_3$ (dark blue). }
    \label{fig:RRate}
\end{figure}

\begin{figure}%[!htbp]
    \centering
    \includegraphics[width=\columnwidth]{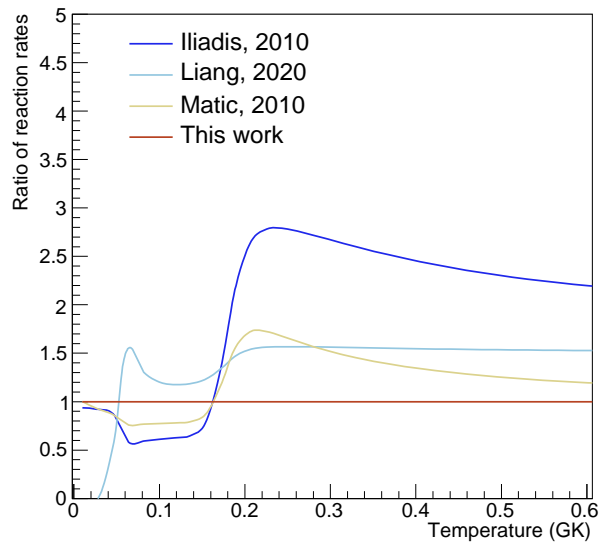}
    \caption{Ratio of reaction rate from different rates reported in JINA's \texttt{REACLIB} database \cite{JINA2010}, compared to the current work.}
    \label{fig:RRateDifference}
\end{figure}

In Fig.~\ref{fig:RRate} each contribution to the reaction rate is plotted for temperatures up to 0.6 GK. As expected, above 0.2 GK, the $3^{+}_{3}$ state dominates the total rate. The effect of the updated resonance parameters are shown in Fig.~\ref{fig:RRateDifference} where the rate from this work is compared to previously determined rates reported in the \texttt{REACLIB} database \cite{JINA2010}.

%%%%%%%%%%%%%%%%%%%%%%%%%%%%%%%%%%%%%%%%%%%%%%%%%%%%%%%%%%%
\section{\label{sec:Conclusion} Conclusion}

The proton decay from the $\mathrm{^{25}Al(d,n)^{26}Si}$ reaction was measured and the total cross section was obtained for the important $3^+_3$ state at 5.92 MeV with higher precision and improved statistics than previously measured by Peplowski {\it et al.} \cite{Peplowski2009}. A CRC calculation, using \verb|FRESCO|~\cite{Fresco}, was used to extract spectroscopic information of the lowest $\ell=0$ proton resonance, taking into account additional information obtained from the isospin mirror nucleus $^{26}$Mg. An updated proton width, $\Gamma_{p} = 2.19(45)$ eV, for the $3^+_3$ state in $^{26}$Si was extracted using the adopted resonance energy 0.4154(8) MeV and the branching ratio from Liang {\it et al.} \cite{Liang2020}. The resulting resonance strength, 26(10) meV, was used to calculated the total reaction rate. Due to the decreased resonance strength of the $3^+_3$ state, this work suggests a reduction in the reaction rate at temperatures above 0.2 GK.

%%%%%%%%%%%%%%%%%%%%%%%%%%%%%%%%%%%%%%%%%%%%%%%%%%%%%%%%%%%
% \newpage
\begin{acknowledgments}

This work is partially supported by the National Science Foundation, under Award Number PHY-2012522. This material is partially based upon work supported by the Department of Energy, National Nuclear Security Administration, under Award Number DE-NA0003841.  The authors acknowledge the support of D. Caussyn, P.A. Barber, B. Schmidt and D. Spingler at the John D. Fox Accelerator Laboratory at Florida State University.

\end{acknowledgments}

\appendix
\bibliography{main.bib}% Produces the bibliography via BibTeX.
\end{document}